\documentclass[aps,prl,twocolumn,groupedaddress,amsmath,amssymb]{revtex4}
\usepackage{graphicx}
\begin{document}

\title{Near-Complete Teleportation of a Superposed Coherent State }
\author{Yong Wook Cheong}
\author{Hyunjae Kim}
\author{Hai-Woong Lee}
\affiliation{Department of Physics, Korea Advanced Institute of
Science and Technology, Daejeon 305-701, Korea}
\date{\today}
\begin{abstract}
The four Bell-type entangled coherent states,
$|{\alpha}\rangle|{-\alpha}\rangle\pm|{-\alpha}\rangle
|{\alpha}\rangle$ and $|{\alpha}\rangle | {\alpha}\rangle \pm
|{-\alpha}\rangle |{-\alpha}\rangle$, can be discriminated with a
high probability using only linear optical means, as long as
$|\alpha|$ is not too small. Based on this observation, we propose
a simple scheme to almost completely teleport a
superposed coherent state. The nonunitary transformation, that is
required to complete the teleportation, can be achieved by
embedding the receiver's field state in a larger Hilbert space
consisting of the field and a single atom and performing a unitary
transformation on this Hilbert space.
\end{abstract}
\pacs{pacs} \maketitle

\newcommand{\bra}[1]{\left<#1\right|}
\newcommand{\ket}[1]{\left|#1\right>}
\newcommand{\abs}[1]{\left|#1\right|}

Since its first proposal\cite{bennett}, a large number of
theoretical and experimental investigations have shown that
teleportation of superposed polarization states\cite{bouwm},
superposed one particle and vacuum states\cite{lee},
and continuous-variable light states\cite{vaid} can be accomplished. The major obstacle to a
demonstration of a complete teleportation for discrete systems with 100\% success
probability is that a clear distinction of four Bell states, a
necessary requirement for a success-guaranteed teleportation, is
not possible with linear optical means\cite{lut}. Although this
obstacle can in principle be overcome by various means such as
exploiting feedback from detectors, employing additional degrees
of freedom or utilizing nonlinear optical
interactions\cite{knill}, its experimental realization
seems difficult\cite{kim}.\\
\indent Quantum teleportation of a superposed coherent
state(known as the Schr\"odinger cat state), a superposition of two
nonorthogonal coherent states with opposite phases, has also been
studied in the past\cite{enk}. Theoretical
investigations have revealed that an interesting feature that
characterizes teleportation of a superposed coherent state
 is that the four ``quasi-Bell states'', four Bell-type
entangled coherent states\cite{sand}, can be distinguished with a
probability approaching unity using only linear optical devices.
The main purpose of this work is to propose a simple scheme that
can perform a near-complete teleportation of a superposed coherent state
 with the success probability and fidelity of nearly 100\%. The
scheme uses linear optics except in the state transformation
process that needs to be carried out in the final stage of the
teleportation, where an additional system, an atom, is brought in
contact with the field and the subsequent atom-field interaction
is utilized.\\
\indent It should be emphasized  that, while the teleportation scheme for
continuous-variable light states(in which a 100\% success probability can
in principle be achieved through ideal homodyne detection\cite{vaid}) is designed
to teleport the quadrature amplitudes of a light field, our scheme takes a discrete
system with the coherent states of opposite phases, $\ket{\alpha}$ and $\ket{-\alpha}$,
as the two basis states for a qubit. An input state for the continuous-variable teleportation
is typically a coherent state, whereas for our teleportation scheme it is a superposed coherent state.\\
\indent The scheme we propose is identical to the standard linear-optical teleportation
scheme, except that Bob's station needs to be equipped with an
additional system, i.e., a single atom trapped in a cavity, to
perform a state transformation, as described later. At the source
station, the entangled coherent state
\begin{equation}
\ket{\Psi}_{AB}=N
(\ket{\alpha}_{A}\ket{-\alpha}_{B}-\ket{-\alpha}_{A}\ket{\alpha}_{B})
\end{equation}
is generated, where A and B refer to the waves that are sent to Alice and Bob, respectively,
and $N(=1/\sqrt{2(1-e^{-4|\alpha|^2})})$ is the
normalization constant. At Alice's station the wave A is combined via a 50/50 beam splitter with
 another wave C that contains an unknown superposed
coherent state to be teleported,
\begin{equation}
\ket{\Psi}_{C}=x\ket{\alpha}_{C}+y\ket{-\alpha}_{C}
\end{equation}
where the unknown coefficients $x$ and $y$ satisfy the
normalization condition $_{C}\langle\Psi\ket{\Psi}_{C}=1$.
The state $\ket{\Psi}_{AB}\ket{\Psi}_{C}$ is transformed, by the
action of the beam splitter, into the state $\ket{\Psi}_{EFB}$,
\begin{eqnarray}
&&\ket{\Psi}_{EFB}\nonumber\\
&&=\frac{N}{2}\{-\ket{0}_E(\ket{\sqrt{2\alpha}}_{F}-\ket{\sqrt{-2\alpha}}_{F})(x\ket{\alpha}_{B}+y\ket{-\alpha}_{B})\nonumber\\
&&-\ket{0}_E(\ket{\sqrt{2\alpha}}_{F}+\ket{\sqrt{-2\alpha}}_{F})(x\ket{\alpha}_{B}-y\ket{-\alpha}_{B})\nonumber\\
&&+(\ket{\sqrt{2\alpha}}_{E}-\ket{\sqrt{-2\alpha}}_{E})\ket{0}_F(x\ket{-\alpha}_{B}+y\ket{\alpha}_{B})\nonumber\\
&&+(\ket{\sqrt{2\alpha}}_{E}+\ket{\sqrt{-2\alpha}}_{E})\ket{0}_F(x\ket{-\alpha}_{B}-y\ket{\alpha}_{B})\}
\end{eqnarray}
where E and F denote the waves that exit through the two output ports of the beam splitter,
which lead to the detectors $D_{E}$ and $D_{F}$, respectively.
Eq.(3) indicates that, as has already been
noted\cite{enk}, the four ``quasi-Bell states'',
$(\ket{\alpha}\ket{-\alpha}\pm\ket{-\alpha}\ket{\alpha})$ and
$(\ket{\alpha}\ket{\alpha}\pm\ket{-\alpha}\ket{-\alpha})$, can be
distinguished by observing which detector, $D_{E}$ or $D_{F}$,
measures an odd or even number of photons and which detects no
photon. The only indistinguishable case is when both detectors
detect no photon, in which case no distinction can be made between
$(\ket{\alpha}\ket{-\alpha}+\ket{-\alpha}\ket{\alpha})$ and
$(\ket{\alpha}\ket{\alpha}+\ket{-\alpha}\ket{-\alpha})$. The
probability that Alice's Bell measurement fails is then given by
$P_F=|_F\bra{0}_E\bra{0}\Psi\rangle_{EFB}|^2=|_C\bra{0}_A\bra{0}\Psi\rangle_{AB}|\Psi\rangle_{C}|^2$
and can easily be calculated to be
\begin{equation}
P_F=P_F(x,y)=\frac{e^{-2|\alpha|^2}}{1+e^{-2|\alpha|^2}}|x+y|^2
\end{equation}
This is negligible for sufficiently large $\abs\alpha$.

After being informed of Alice's Bell measurement result, Bob needs
to perform an appropriate transformation on the field state B to
complete the teleportation. The difficulty arises when
either of the two detectors
measures a nonzero even number of photons, because then a nonunitary
transformation is needed. This nonunitary transformation can be
accomplished approximately by applying a unitary transformation
that displaces the state by an appropriate amount\cite{jeong02},
or probabilistically by performing teleportation  repeatedly with
entangled ancilla photons until the desired transformation is
reached\cite{ralph}. We show below that one can perform the
required nonunitary transformation successfully with a high
probability by bringing an additional system, a single atom trapped in a cavity, in
contact with the field to be transformed and utilizing the
interaction between the field and the atom.

\begin{figure}[t]
\centering
\includegraphics[width=0.45\textwidth]{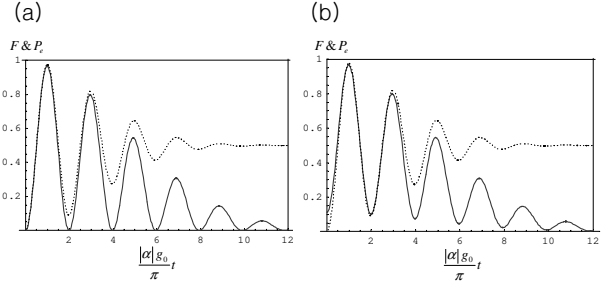}
\caption{\label{fig1} The fidelity $F$(solid curve) of Eq.(5) and
the excitation probability $P_e$(dotted curve)as a function of
time for the case $\alpha=5$ and
(a)$x=y=\frac{1}{\sqrt{2(1+e^{-50})}}$
,(b)$x=\sqrt{2}y=\sqrt{\frac{2}{3+2\sqrt{2}e^{-50}}}$.}
\end{figure}
Let us consider the case when Detector $D_E$ detects no photon and
$D_F$ measures a nonzero even number of photons. The
transformation needed is $M(x\ket{\alpha}-y\ket{-\alpha})
\Rightarrow \left( x\ket{\alpha}+y\ket{-\alpha} \right)$, where
$M$ is the normalization constant($M=\frac{1}{\sqrt{2(|x|^2+|y|^2)-1}}$).
In order to achieve this
transformation, we allow the field to interact with a single
two-level atom prepared in its lower level $\ket{g}$. We assume
that the atomic transition between the upper level $\ket{e}$ and
the lower level $\ket{g}$ is resonant with the field frequency.
The state of the system, atom+field, at the initial time, $t=0$,
is $\ket{\psi(t=0)}=M\ket{g}(x\ket{\alpha}-y\ket{-\alpha})$.
The state at a later time $t>0$ is given, under the rotating wave
approximation, by the solution of the Jaynes-Cummings model, provided that
 spontaneous emission and the cavity decay can be neglected.
We then obtain, for the fidelity of the field state at time $t$ with respect
to the desired state ($x\ket{\alpha}+y\ket{-\alpha}$),
\begin{eqnarray}
F(x,y)=(x^{*}\bra{\alpha}+y^{*}\bra{-\alpha}) \rho
(x\ket{\alpha}+y\ket{-\alpha}) =|M|^2e^{-2|\alpha|^2} \nonumber\\
\times \{
|\sum_{n=0}^{\infty}{\frac{|\alpha|^{2n}}{n!}(|x|^2-|y|^2+2(-1)^ni
\mbox{Im} (xy^{*})
)\cos(\frac{\sqrt{n}g_{0} t}{2})|^2} \nonumber\\
+|\sum_{n=0}^{\infty} {\frac{|\alpha|^{2n+1}}{\sqrt{n!(n+1)!}}
|x+(-1)^{n}y|^2 \sin{(\frac{\sqrt{n+1}g_{0}t}{2} )}|^2} \}
\end{eqnarray}
where $\rho$ is the reduced density operator of the field at time t, and
$g_{0}$ is the single-photon Rabi frequency.
The
fidelity $F(x,y)$ depends upon $x$ and $y$ as well as $\alpha$ and
$t$. In Figs.~1(a) and 1(b), we show the fidelity $F$ we computed
as a function of time for the case $\alpha=5$ and for
$x=y=\frac{1}{\sqrt{2(1+e^{-50})}}\simeq \frac{1}{\sqrt{2}}$, and
$x=\sqrt{2}y=\sqrt{\frac{2}{3+2\sqrt{2}e^{-50}}}\simeq
\sqrt{\frac{2}{3}}$, along with the probability $P_e$ at time $t$
that the atom is found in its upper level $\ket{e}$.
\begin{figure}[t]
\centering
\includegraphics[width=0.35\textwidth]{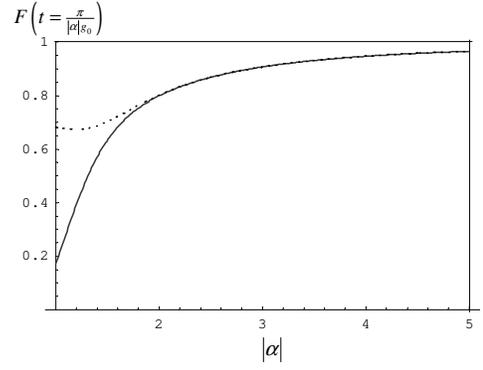}
\caption{ \label{fig2} The fidelity
$F(t=\frac{\pi}{\abs{\alpha}g_{0}})$ as a function of
$\abs{\alpha}$ for the case
$x=y=\frac{1}{\sqrt{2(1+e^{-2|\alpha|^2})}}$(solid curve) and for
the case $x=-y=\frac{1}{\sqrt{2(1-e^{-2|\alpha|^2})}}$(dotted
curve) }
\end{figure}
 It is seen that $F$ and $P_e$ show
similar temporal behavior. The time at which $F$ takes on the
largest value is given roughly by
$t\simeq\frac{\pi}{\abs{\alpha}g_{0}}$, as long as $\abs{\alpha}$
is not too small($\abs{\alpha}\gtrsim 3$). Fig.2 shows
$F(t=\frac{\pi}{\abs{\alpha}g_{0}})$, the fidelity $F$ at
$t=\frac{\pi}{\abs{\alpha}g_{0}}$, as a function of $\alpha$ for
the case $x=y=\frac{1}{\sqrt{2(1+e^{-2|\alpha|^2})}}$ (when
$\ket{\Psi}_c$ is the even coherent state) and for the case
$x=-y=\frac{1}{\sqrt{2(1-e^{-2|\alpha|^2})}}$ (when $\ket{\Psi}_c$
is the odd coherent state). It is seen that , as $\abs{\alpha}$
becomes larger, the fidelity at $t=\frac{\pi}{\abs{\alpha}g_{0}}$
becomes close to unity. We note that the fidelity
$F(t=\frac{\pi}{\abs{\alpha}g_{0}})$ varies slowly with respect
to $x$ and $y$ for $\abs{\alpha}\gtrsim 1$. For example , the plot
for $F(t=\frac{\pi}{\abs{\alpha}g_{0}})$ for the case
$x=\sqrt{2}y=\sqrt{\frac{2}{3+2\sqrt{2}e^{-50}}}$ and for the case
$x=-\sqrt{2}y=\sqrt{\frac{2}{3-2\sqrt{2}e^{-50}}}$ is almost
indistinguishable from Fig.2. It should be remarked that, when we
fix the values of $\alpha$, $x$ and $y$ and follow the time
variation of the fidelity F, the maximum value $F_{max}$ occurs at
the time very close to but not exactly equal to
$\frac{\pi}{\abs{\alpha}g_{0}}$. Thus, the actual maximum value
$F_{max}$ of the fidelity is slightly greater than the value
$F(t=\frac{\pi}{\abs{\alpha}g_{0}})$ that Fig.2 indicates. The
difference, however, is negligibly small if $\abs{\alpha}\gtrsim
3$. One can conclude that, as long as $\abs{\alpha}$ is not too
small($\abs{\alpha}\gtrsim 3$), the transformation of the field
state from $M(x\ket{\alpha}-y\ket{-\alpha})$ to
$(x\ket{\alpha}+y\ket{-\alpha})$ can be achieved with high
fidelity by allowing the field to interact with a single two-level
atom in its lower level and waiting for a time of
$\frac{\pi}{\abs{\alpha}g_{0}}$(i.e. waiting until the atom is
excited). At this instant, if one wishes, one can measure the
state of the atom, decoupling the atom from the field, and confirm
that the atom has indeed been excited.

When Detector $D_E$ measures a nonzero even number of photons and
Detector $D_F$ detects no photon,
Bob needs to first apply the unitary transformation
$(-1)^{a^{\dag} a}$ to the field state and then allow the field
to interact with a single atom as described above.

We remark that, as our analysis of the atom-field interaction neglects
spontaneous emission and the cavity decay, the successful  operation of our scheme requires
a setting in which coherent atom-field interaction dominates dissipation.
Specifically, it requires 
$\sqrt{\bar{n}}g_{0} \gg \gamma$, and $\sqrt{\bar{n}}g_{0} \gg \bar{n} \kappa $,
where $\bar{n}(=|\alpha|^2)$ is the average photon number in the field,
$\gamma$ the spontaneous decay rate of the atom,
and $\kappa$ the cavity field decay rate.
We bring attention to the factor $\bar{n}$ on the right hand side of the second inequality,
which arises because the superposed coherent state interacting with an atom in a cavity
decoheres fast into a statistical mixture in time $\sim 1/\bar{n} \kappa$ \cite{walls,bru}.
Because of this factor, strong coherent fields cannot be used in our scheme.
As an example, let us take the experimental parameters, $\frac{g_{0}}{2\pi}=47KHz$, $\tau_{sp}=\frac{1}{\gamma}=30ms$,
$\tau_{c}=\frac{1}{\kappa}=1ms$, quoted in a recent cavity
quantum electrodynamics(qed) experiment with circular Rydberg states of rubidium atoms in a millimeter wave
superconducting high-finesse cavity\cite{rai}.
The first inequality is easily satisfied, but the second inequality requires $\sqrt{\bar{n}} \ll \frac{g_{0}}{\kappa} \approx 300$,
i.e., $\bar{n} \ll 10^5$.
If we takes  $\frac{g_{0}}{2\pi}=32MHz$, $\frac{\gamma}{2\pi}=2.6MHz$, $\frac{\kappa}{2\pi}=4MHz$ from an experiment
\cite{ye} conducted in the near-infrared regime with cesium atoms trapped in a far-off-resonance trap,
the first inequality requires $\sqrt{\bar{n}} \gg  \frac{\gamma}{g_{0}} \approx 0.081$, i.e., $\bar{n} \gg 0.066$,
and the second requires $\sqrt{\bar{n}} \ll \frac{g_{0}}{\kappa} =8$, i.e.,$\bar{n} \ll 64$.

We now wish to calculate the average fidelity of the entire
teleportation process. The average fidelity $F_{ave}$ can be
defined as $F_{ave}=\sum_{i=1}^{5}\overline{P_i(x,y)F_i(x,y)}$,
where $P_1,P_2,P_3,P_4$ and $P_5$ represent, respectively, the
probability that the number of photons measured by
detectors($D_E$,$D_F$) are ($0$,odd), (odd,$0$),
($0$,nonzero-even), (nonzero-even,$0$), and ($0$,$0$); $F_i$
denotes the fidelity of the field state B with respect to the
original state of Eq.(2), obtained after an appropriate
transformation performed according to the result of Alice's Bell
measurement; and the bar on the right-hand-side
indicates averaging over the unknown coefficients $x$ and $y$. It
is clear that $P_1=P_2=\frac{1}{4}$,$P_5=P_F(x,y)$(see Eq.(4)),
$P_3=P_4=\frac{1}{4}-\frac{1}{2}P_F(x,y)$, $F_1=F_2=1$, and
$F_3=F_4=F_{max} (x,y)$, where the actual maximum value $F_{max}
(x,y)$ of the fidelity of Eq.(5) can be replaced by $F(x,y)$ at
time $t= \frac{\pi}{\abs{\alpha} g_{0}}$ if $\abs{\alpha}$ is
sufficiently large. We can also calculate easily $F_5$ and obtain
$F_5=\frac{1-e^{-2|\alpha|^2}}{2}\abs{x-y}^2$. In order to perform
averaging over $x$ and $y$, it is convenient to express the state
$\ket{\Psi}_c$ of Eq.(2) in orthonormal bases. We
choose to express it in terms of even and odd coherent states
$\ket{\alpha_e}$ and $\ket{\alpha_o}$\cite{dod} as
$\ket{\psi}_{C}=\sin{\frac{\theta}{2}}\ket{\alpha_e}_{C}+\cos{\frac{\theta}{2}}e^{i\phi}\ket{\alpha_o}_{C}$.
The probabilities $P_{i}$ and the fidelities $F_{i}$ can be
expressed as a function of $\theta$ and $\phi$, $P_{i}(\theta,
\phi) $ and $F_{i}(\theta, \phi)$, and the averaging can then be
performed according to
$F_{ave}=\int_{0}^{2\pi}d\phi \int_{0}^{\pi}\sin{\theta}d\theta
{\sum_{i=1}^{5}P_i(\theta ,\phi)F_i(\theta,\phi)}$.

In Fig.3, we show the average fidelity $F_{ave}$, as a function of
$\alpha$, that we computed using the Monte-Carlo method, where the
solid and dotted curves are obtained by using the actual maximum
value $F_{max}$ and $F(t=\frac{\pi}{\abs{\alpha}g_{0}})$,
respectively, for the value of the fidelity $F_3=F_4=F(x,y)$.
\begin{figure}[t]
\includegraphics[width=0.35\textwidth]{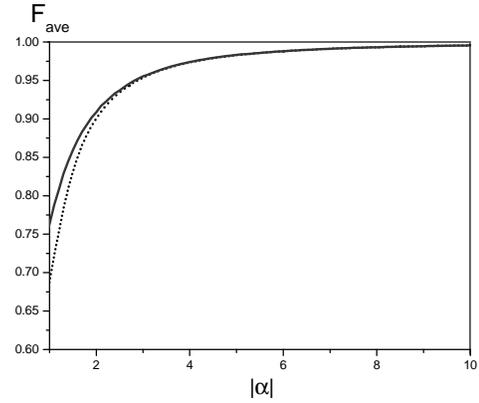}
\caption{\label{fig3} The average fidelity $F_{ave}$ as a function
of $\abs\alpha$. The solid and dotted curves represent $F_{ave}$
computed using, for the value of $F_3=F_4$, the actual maximum
value $F_{max}$ and the value of the fidelity at
$t=\frac{\pi}{\abs{\alpha}g_{0}}$, respectively.}
\end{figure}
Compared with the standard linear optical scheme of teleporting a
superposed polarization state, where a beam splitter and two
polarizing beam splitters are used for Bell-state measurements, in
which case the average fidelity is given by $\frac{5}{6}$(When the
Bell-state measurement is successful, which occurs with the probability of
$1/2$, the fidelity is  $1$. When it is not, which also occurs with the probability of
$1/2$, the maximum possible fidelity averaged over the unknown coefficients after a
bit-flip operation by Bob is $2/3$, being equivalent to the case when Alice and Bob
are connected only by a classical channel\cite{pope}. The average fidelity is thus given by
$\frac{1}{2}\times 1 +\frac{1}{2} \times \frac{2}{3}=\frac{5}{6}$), the
present case yields a higher average fidelity for
$\abs{\alpha}\gtrsim 1.33$. If $\abs{\alpha}$ is not too small,
the average fidelity is close to unity(for example,
$F_{ave}=0.955$ at $\abs{\alpha}=3$ ), and for reasonably large
$\abs{\alpha}$, $F_{ave}$ can be regarded to be practically equal
to 1.

Finally, we remark on the issue concerning experimental
realization of our proposed scheme. The scheme requires generation
and manipulation of the entangled coherent state,
and
photodetection that distinguishes between even and odd photons,
all of which are highly demanding experimentally. There have,
however, been suggestions and proposals that would help to
overcome the difficulties. The entangled coherent state of Eq.(1)
can be generated by illuminating a 50/50 beam splitter with a
superposed coherent state
$\ket{\sqrt{2}\alpha}-\ket{-\sqrt{2}\alpha}$ through one of the
input ports. The superposed coherent state can in turn be
generated by exploiting nonlinear interaction between a coherent state field and atoms\cite{bru,yurke},
or by means of a conditional
measurement on a beam splitter\cite{dakna}.
Discriminating even to odd photons requires in
principle detectors with single photon resolution that can
distinguish between $n$ and $(n+1)$ photons. Visible light photon
counters have recently been constructed that can distinguish
between no photon and a single photon with the quantum efficiency
exceeding $70\%$\cite{kwiat02}, and that can distinguish between a
single photon and two photons with the quantum efficiency of
$47\%$\cite{jkim}.  Using such counters in an arrangement of
detector cascades\cite{song} or N-ports \cite{reck}, it is in
principle possible to distinguish between  $n$ and $(n+1)$
photons. It has also been suggested that  $n$ and $(n+1)$ photons
can be distinguished by utilizing homodyne detection looking at
the imaginary quadrature \cite{ralph}, or by coupling the field to
a two-level atom through nonlinear interaction \cite{yurke02}.

 In conclusion we have shown that a near-complete teleportation
of a superposed coherent state is possible using only a
linear-optical scheme and an atom-field interaction. The average
fidelity of the proposed scheme exceeds that of the standard
linear-optical scheme for teleporting a superposed polarization
state, as long as $\abs{\alpha}\gtrsim 1.33$.

ACKNOWLEDGMENT

This research was supported by Korea Research Foundation
Grant(KRF-2002-070-C00029). We wish to thank Drs. J. Kim and B.A.
Nguyen of KIAS and Professor M.S. Kim and Dr. H. Jeong of Queens
University, Belfast for helpful discussions.

\end{document}